\documentstyle[pre,aps,multicol,epsfig,amssymb]{revtex}
\begin{document}

\title{Many-body interactions and correlations in coarse-grained
descriptions of  polymer solutions}

\author {P.G. Bolhuis\protect\footnote{email: bolhuis@its.chem.uva.nl}, 
A.A. Louis\protect\footnote{email: ardlouis@theor.ch.cam.ac.uk}, and J.P. Hansen}
 \address{Department of Chemistry, Lensfield Rd, Cambridge CB2 1EW,
 UK} \date{\today} \maketitle
\begin{abstract}
\noindent 
We calculate the two, three, four, and five-body (state independent)
effective potentials between the centers of mass (CM) of self avoiding
walk polymers by Monte-Carlo simulations.  For full overlap, these
coarse-grained $n$-body interactions oscillate in sign as $(-1)^n$,
and decrease in absolute magnitude with increasing $n$.  We find
semi-quantitative agreement with a scaling theory, and use this to
discuss how the coarse-grained free energy converges when expanded to
arbitrary order in the many-body potentials. We also derive effective
{\em density dependent} 2-body potentials which exactly reproduce the
pair-correlations between the CM of the self avoiding walk polymers.
The density dependence of these pair potentials can be largely
understood from the effects of the {\em density independent} 3-body
potential.  Triplet correlations between the CM of the polymers are
surprisingly well, but not exactly, described by our coarse-grained
effective pair potential picture.  In fact, we demonstrate that a
pair-potential cannot simultaneously reproduce the two and three body
correlations in a system with many-body interactions.  However, the
deviations that do occur in our system are very small, and can be
explained by the direct influence of 3-body potentials.
\end{abstract} 
\begin{multicols}{2}

\section{Introduction}
An efficient statistical description of condensed matter systems and
materials almost invariably involves some degree of coarse-graining,
whereby a large fraction of the initial microscopic degrees of freedom
are traced out, leaving a much reduced space of variables associated
with the composite entities or pseudo-particles.  The latter are then
coupled via effective interactions which result from the partial
averaging over the initial microscopic degrees of freedom.  The
reduction of the initial multicomponent system to a coarse-grained
system with a substantially smaller set of composite particles implies
that the resulting effective interactions may involve three-body and
higher-order contributions, even if the original multicomponent system
involved only pair-wise additive forces, like Coulombic interactions.
Alternatively, one may wish to retain the simplicity of pair-wise
additivity of the effective interactions, but the price to pay is that
such effective pair potentials are then state-dependent, e.g.\ are
functions of the temperature and/or density.  The reason for this is
that the effective interaction energy is a free energy associated with
the averaged-out degrees of freedom, which generally has an entropic
component.

There are many examples of the coarse-graining procedure which has
just been outlined.  In molecular systems the forces between nuclei
result from gradients of the electronic ground state energy surface
which depends parametrically on the nuclear coordinates, and adjusts
adiabatically to the slow motion of the latter within the
Born-Oppenheimer approximation.  This scenario is mimicked, at least
at the level of valence electrons, in ``ab initio'' Molecular Dynamics
simulations pioneered by Car and Parrinello\cite{Car85}.  However in
situations where no strong covalent or hydrogen bonding is present,
the more common route is to represent the total ground state
electronic energy surface by a sum of one-body, two-body and higher
order terms.  The one-body contribution is the sum of the ground state
energies associated with individual, isolated chemical entities
(atoms, ions, or molecules).  The two-body term is made up of the sum
of pair potentials acting between molecules and higher order terms
correspond to isolated clusters of three or more molecules.  The sum
of ground state energies of individual molecules does not contribute
to the forces between them, and can hence be ignored in the
description of collective equilibrium or transport processes which do
not involve chemical reactions.  In the simplest case of rare-gas
atoms, the pair-wise interactions would include overlap repulsion at
short range and dispersion forces at long range, while triplet
interactions would include, among others, the Axilrod-Teller triple
dipole dispersion potential\cite{Axil43}, which contributes very
significantly to the thermodynamic and transport properties of the
heavier rare gases in their condensed states\cite{Bark71,Leve88}.  The
effect of the higher order interactions can be approximately
incorporated into an effective pair potential, which differs from the
bare pair potential, valid for an isolated pair of molecules, and
becomes density-dependent\cite{Reat87,Atta92}. 

Similarly in metals an effective interaction between ions may be
determined by tracing out the conduction electrons, using perturbation
theory or response theory\cite{Ashc78,Carl90}.  Treating the
ion-electron coupling to lowest order (linear response) leads to a
structure-independent volume term and to a pair-wise screened
effective potential between ``dressed'' ions or pseudo-atoms, which
both depend on the macroscopic conduction electron density.  The
two-body level is generally sufficient for alkali and other simple
metals, in part because of a quantum interference effect which
strongly decreases the magnitude of the higher order response
terms\cite{Loui98}.  For multivalent and transition metals many-body
effective interactions can no longer be neglected\cite{Carl90}, and a
full ``ab initio'' treatment may be necessary\cite{Madd93}.

Coarse-graining becomes crucial in the highly asymmetric systems of
soft matter, involving macromolecules or colloidal particles, as well
as molecular scale entities, like solvent molecules or ions.  The
latter are traced out to derive effective interactions between the
electric double-layers associated with charged surfaces (colloids,
membranes etc...)\cite{Hans00}.  The microscopic ions play a role
similar to valence electrons in metals, but with the important
difference that quantum degeneracy effects are absent, and that finite
temperature entropic effects control the width of the electric
double-layers, with a resulting density and temperature dependence of
the effective interactions between the mesoscopic colloidal particles.
A good example of a such a pair potential is provided by the classic
Derjaguin-Landau-Verwey-Overbeek (DLVO) effective pair potential
between spherical charged colloidal particles\cite{Hans00}.
Three-body interactions can be derived in similar
fashion\cite{Lowe98}.  However, the averaging over the microion
degrees of freedom also leads to a structure-independent, but
state-dependent volume term, part of which is associated with the
self-energy of individual double-layers, and which is reminiscent of
the volume term in metals\cite{Grim91,Roij97}; this term has a
profound effect on the phase diagram of charge-stabilized
colloids\cite{Roij97,Roij99}.

Another important class of effective interactions of entropic origin,
which follow from averaging over the configurations of non-adsorbing
polymers and small colloidal particles, are the depletion forces which
have received much renewed interest in recent
years\cite{Dijk98,Liko01}.  Depletion pair potentials depend
strongly on the concentration of the depletant; recent attempts have
been made to compute the three-body interactions from simulations or
density functional theory\cite{Melc00}.  Volume terms arising from the
depletion potential picture may have an important effect on the
osmotic equation of state, but they are not expected to influence the
phase behavior in these uncharged systems\cite{Dijk98}.

The present paper is concerned with a coarse-grained description of
dilute and semi-dilute solutions of polymers in good solvent.  Whereas
tracing out the microscopic ions in a charged colloidal suspension has
many analogies with the liquid metal problem, the coarse-graining of
neutral polymers, achieved by integrating out the internal monomeric
degrees of freedom, resembles more closely the case of effective
potentials between neutral atoms and molecules, obtained by tracing
out the internal electronic degrees of freedom.  The basic idea
explored here, which goes back to Flory and Krigbaum\cite{Flor50}, is
to represent a set of polymers, each made up of $L$ monomers or
segments, as single particles, interacting with each other through an
effective interaction between their centers of mass (CM).  The
important point, realized by Grosberg {\em et al.\ }\cite{Gros82}, is
that the effective pair interaction remains finite, even for
infinitely long polymers.  Monte Carlo simulations\cite{Daut94} and
renormalization group calculations\cite{Krug89} show that for two
isolated non-intersecting polymer coils, the effective potential
between their CM's is of order $2 k_B T$ in the scaling limit, i.e.\
for $L$ going to infinity, while the range of the interaction is of
the order of the radius of gyration, $R_g$, of the polymers.  Recently
we have extended this investigation by simulating large systems of
self-avoiding walk (SAW) polymers at {\em finite}
concentration\cite{Loui00,Bolh01a}.  The resulting CM pair
distribution function $g(r)$ was then inverted to yield a
concentration-dependent effective pair potential $v(r;\rho)$, where
$\rho$ is the number of polymer coils per unit volume.  Although
$v(r;\rho)$ was not found to change dramatically with $\rho$, the
$\rho$-dependence of $v$ is very significant for the accurate
determination of the osmotic properties of dilute and semi-dilute
polymer solutions.

In this paper we adopt a somewhat different point of view, by
determining state-independent effective pair, triplet, quadruplet and
quintuplet interactions; these  many-body interactions are determined
by successively considering clusters of 2, 3, 4, and 5 SAW polymer
chains, determining the corresponding n-body distribution functions,
from which an effective n-body potential is derived.  Triplet
interactions between the cores of star-polymers have recently been
determined in a similar way\cite{vonF00}.  The next step taken in this
paper is to relate the low-density (state-independent) pair and triplet
interactions to the density-dependent effective interactions
determined in our earlier work\cite{Loui00,Bolh01a}.  A final section
will be devoted to an analysis of three-body correlations as measured
by bond-angle distributions and deviations from the Kirkwood
superposition approximation\cite{Kirk52}.  The convergence of the
series of n-body interactions is assessed on the basis of scaling
arguments in the Appendix.

\section{Simulation Models and Methods}
\label{sec:model}
Many properties of polymers in a good solvent are well described by
models which ignore all microscopic details of the intermolecular
interactions, except their excluded volume.  For that reason polymers
are often modeled as self avoiding walks (SAW) on a
lattice\cite{Flor53,deGe79,Doi95}, a model lending
itself well to efficient computer simulations.  We consider the
situation of $N$ athermal SAW chains of length $L$ on a simple cubic
lattice of $M$ sites.  The bead or segment concentration is given by
$c=NL/M$, while the polymer chain concentration is given by
$\rho=N/M$. The polymers are characterized by their radius of gyration
$R_g$ which, for an isolated polymer, scales as $R_g \sim L^{\nu}$,
where $\nu\approx 0.59$ is the Flory
exponent\cite{Flor53,deGe79,Doi95}.  We also define an overlap
concentration $\rho^* = 1/ \frac{4}{3} \pi R_g^3$ at which there is on
average one polymer per sphere of radius $R_g$.  Solutions with
$\rho/\rho^* < 1$ are called dilute, while solutions with $\rho/\rho^*
> 1$ {\em and} $c \ll 1$ are called semi-dilute.  When the monomer
density $c$ becomes appreciable, the solution moves from the
semi-dilute to the melt regime.  In this paper we will focus on
densities $\rho/\rho^* \lesssim 2$, i.e.\ the dilute regime and the
beginning of the semi-dilute regime.

When modelling the semi-dilute regime, it is important to take
 sufficiently long polymer chains.  The first reason is that for
 studying the semi-dilute regime one needs a large polymer density
 $\rho$ together with a low monomer density $c$.  We found earlier
 that the monomer density $c^*$ at the overlap concentration $\rho^*$
 scales roughly like\cite{Bolh01a} $c^* \approx 4 L^{-0.8}$ for SAW
 polymers on a simple cubic lattice.  Throughout this paper we use
 polymers of length $L=500$ for which $c\approx 0.05$ at
 $\rho/\rho^*=2$, so that we are still clearly in the semi-dilute
 regime.  In contrast, for $L=100$ the monomer density is $c \approx
 0.2$ at $\rho/\rho^*=2$, suggesting that a meaningful semi-dilute
 regime does not exist for such short polymers.

The second reason for using long polymer chains is that we want to
study properties --- particularly the effective potentials between
polymer chains --- in the scaling regime, where all length dependence
is completely captured by $R_g$.  In a previous paper\cite{Bolh01a} we
established that two properties of the effective potential relevant to
thermodynamics, namely the second-virial coefficient $B_2$ between two
polymers and the effective pair-potential between CM's, multiplied by
the square of the CM distance, $r^2v(r)$, are very close to the scaling
limit for $L=500$ polymers, the length we will use in this paper.

Simulations were done with the Monte Carlo pivot
algorithm~\cite{Daut94,pivot} combined with simple translational
moves. For concentrations $\rho>\rho^*$ we also use Configurational
Bias Monte Carlo algorithms~\cite{frenkelbook,dijkstra}.  For $L=500$
polymers we find that the radius of gyration of an isolated coil is
$R_g =16.50 \pm 0.03$.  We used a simulation box of size $M= (240)^3$,
and varied the number of polymers from $N=2$ for the 2-body
calculations to $N=6400$, which corresponds to $\rho/\rho^*=8.9$.

Since we are dealing with athermal chains consisting of monomers
interacting only via hard-core repulsion, we set the reciprocal
temperature $\beta=1/k_B T=1$ throughout this paper.

\section{Density independent many-body interactions}

\subsection{Expanding the coarse-grained free energy in a series of many-body
interactions}

\begin{figure}
\begin{center}
\epsfig{figure=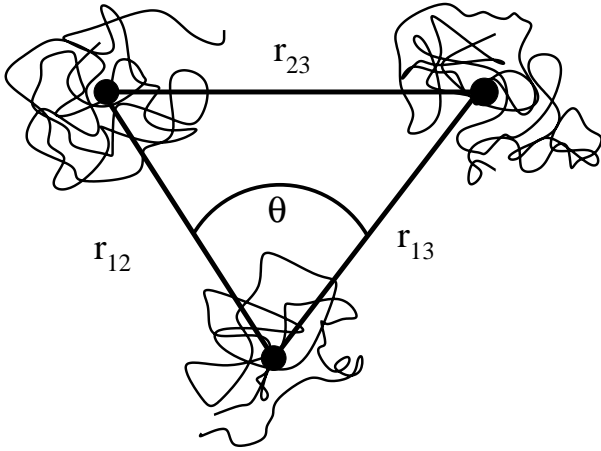,height=8cm,angle=-90}
\caption{\label{fig:3-polymers} The three variables $\{ r_{12},
r_{13}, r_{23}\}$ which characterize a triplet configuration.  The angle
$\theta$ between the vectors ${\bf r}_{12}$ and ${\bf r}_{13}$ is
related to the distance $r_{23}$ by $r_{23}^2 = r_{12}^2 + r_{13}^2 -
2 r_{12} r_{13} \cos \theta$.  }
\end{center}
\end{figure}

Following the discussion in\cite{Loui01a}, the Helmholtz free energy
${\cal F}$ of a set of $N$ polymers of length $L$ with their centers
of mass (CM) distributed according to the set of coordinates $\{{\bf
r}_i\}$, in a volume $V$, can be written as the following expansion:
\begin{eqnarray}\label{eqIII.1}
{\cal F}(N,V,\{{\bf r}_i\})& =& {\cal F}^{(0)}(N,V) + \sum_{i_1<i_2}^{N}
w^{(2)}({\bf r}_{i_1},{\bf r}_{i_2},) \\ \nonumber &+ & \sum^{N}_{i_1<i_2<i_3} w^{(3)}({\bf
r}_{i_1},{\bf r}_{i_2},{\bf r}_{i_3}) + \ldots \\ \nonumber \ldots & + &
\sum^{N}_{i_1 < \ldots < i_n} w^{(N)}({\bf r}_{i_1},{\bf
r}_{i_2}\ldots {\bf r}_{i_N})
\end{eqnarray}
 In the scaling limit, each term in the series is independent of $L$
as long as the n-tuple CM coordinates $\{ {\bf r}_{i_1},{\bf
r}_{i_2}\ldots {\bf r}_{i_n} \}$ are expressed in units of $R_g$, the
radius of gyration at zero density.  Note that this coarse-grained
free energy includes an implicit statistical average over all the
monomeric degrees of freedom for a fixed configuration $\{{\bf r}_i\}$
of the CM.  The full free energy of the underlying polymer system can
be calculated as follows:
\begin{equation}\label{eqIII.1b}
F(N,V) = -\ln \sum_{\{{\bf r}_i\}} \exp \left[-{\cal F}(N,V,\{{\bf
r}_i\})\right]
\end{equation}
so that Eq.~(\ref{eqIII.1}) can be viewed as an expansion of the
effective interaction between the CM in terms of (entropic) many-body
interactions.  ${\cal F}^{(0)}(N,V)$ is the so-called volume term, the
contribution to the free energy which is independent of the
configuration $\{{\bf r}_i\}$\cite{Roij97}.  Here it includes the
free energy of a single isolated polymer which is independent of the
position of its CM in a homogeneous solution; translational invariance
also implies that there is no one-body term in the expansion.  Each
subsequent term $w^{(n)}({\bf r}_{i_1},{\bf r}_{i_2}\ldots {\bf
r}_{i_n})$ is defined as the free energy of $n$ polymers with their CM
positions at $\{ {\bf r}_{i_1},{\bf r}_{i_2}\ldots {\bf r}_{i_n} \}$
minus the contributions of all lower order terms.  In other words, it
is the contribution to the free energy of $n$ polymers which is not
included in the sum of all lower order terms. For instance, the 2-body
term $w^{(2)}(r_{ij})$ can be defined as the difference between the
coarse-grained free energy ${\cal F}$ for two particles with their CM
distance held at $r_{ij} = |{\bf r_i}-{\bf r_j}|$, and the free energy
of the same two polymers when they are infinitely far apart.  Here we
use the translational and rotational invariance of a homogeneous
system to reduce the number of degrees of freedom.  Similarly, the
three-body term for a given triplet configuration $\{ {\bf r}_i,{\bf
r}_j,{\bf r}_k \}$ can be written in terms of only 3 variables (see
Fig.~\ref{fig:3-polymers}) as
\begin{eqnarray}\label{eqIII.2}
w^{(3)}( r_{ij},r_{jk},r_{ki}) & = & {\cal F}( {N=3},V, 
r_{ij},r_{jk},r_{ki}) \nonumber \\ - {\cal F}^{(0)}( {N=3} ,V) & -&
w^{(2)}(r_{ij}) - w^{(2)}(r_{ik}) - w^{(2)}(r_{jk}).
\end{eqnarray} 
In other words, it is that part of the effective interaction between
three polymers which cannot be described by volume and pair
interaction terms alone.  In principle, this procedure may be
continued until, for a system with $N$ polymers, the $N$th term
determines the total coarse-grained free energy.  In practice, this
approach is not feasible because the number of $n$-tuple coordinates
increases rapidly with $n$, as does the complexity of each higher
order term, so that the series in Eq.~(\ref{eqIII.1}) quickly becomes
intractable.  Instead, one hopes to show that the series converges
fast enough that only a few low order terms are needed to obtain a
desired accuracy.  We now turn to the derivation of these density
independent potentials for our system of SAW polymers.

\subsection{2-body interactions}

There is a general relationship between the $\rho \rightarrow 0$ limit
of $n$-body correlation functions and the $n$-body
potential\cite{Loui01a}.  For the 2-body case this reduces to
\begin{equation}\label{eqIII.3}
\lim_{\rho \rightarrow 0} g_2(r) = \exp \left[ -w^{(2)}(r) \right],
\end{equation}
where $g_2(r)$ is the pair distribution function.  Although this
definition resembles that of the potential of mean force (PMF),
usually defined as $ w^{PMF}(r) = -\ln[g_2(r)]$ for any density, they
are only equivalent in the limit of infinite dilution.  Strictly
speaking, the PMF is not a potential but simply a restatement of the
pair-correlations. Consider the simplest case, namely a system with no
higher-order $(n>2)$ interactions. If one were to use a finite density
PMF as a pair potential at that same finite density, the resulting
pair correlations would not be those of the system used to derive the
PMF.  In contrast, the potential defined in Eq.~(\ref{eqIII.3}) is the
correct pair potential which would exactly reproduce the pair
correlations of the original system.

In our simulations we calculate $w^{(2)}(r)$ from the logarithm of the
overlap probability as a function of CM distance (c.f.\
Eq.~\ref{eqIII.3}).  Although the arguments above were made for a
free energy in a continuous space, they easily carry over for the
lattice model we simulate.  In fact, the CM lives on a finer
grid than the original SAW lattice polymers, so that our results are
already closer to the continuum limit.  The overlap probability is
determined by sampling the configurations of two polymers infinitely
far apart with the pivot algorithm, and, after every 1000 pivot moves,
searching for any monomer overlaps as a function of the CM distance.
The effective potential calculated in this manner has a near Gaussian
shape with a value at full overlap of $w^{(2)}(0)=1.88 \pm 0.01$ for
our $L=500$ polymers, very close to the scaling limit estimate 
$w^{(2)}(0)=1.80 \pm 0.05$\cite{Bolh01a}, and a range of the order of
$R_g$\cite{Loui00,Bolh01a}, as shown in Fig.~\ref{fig:v34}.  This
picture agrees with earlier RG\cite{Krug89} and
simulation\cite{Daut94} studies.  Note that in the scaling limit the
potentials depend only on $R_g$, so that the free energy cost of
completely overlapping the CM of two polymers is independent of their
length $L$.  That this free energy cost at complete overlap should
depend weakly on polymer length follows from their fractal
nature\cite{Flor50}, but more sophisticated scaling theory
arguments\cite{Gros82} are needed to prove that $w^{(2)}(0) \propto
L^{0}$\cite{Loui00,Bolh01a}.

\subsection{3-body interactions}

Just as the 2-body interactions follow from the low density limit of
the pair correlations, so also the 3-body or triplet interactions
between the CM of three polymers can be derived by taking the
low-density limit of the 3-body distribution function
\begin{equation}\label{eq:g3}
g_3({\bf r}_1,{\bf r}_2,{\bf r}_3) = \frac{1}{Z} \int ... \int e^{-
{\cal F}(\{ {\bf r}_i \})} d{\bf r}_4 .... d{\bf r}_N,
\end{equation}
where $ {\cal F}(\{ {\bf r}_i \})$ is the coarse-grained
free energy~(\ref{eqIII.1}) and and $Z = \exp[-F]$ is the
configurational integral for the CM coordinates defined in
Eq.~(\ref{eqIII.1b}) the
effective interactions already include an average over monomeric
degrees of freedom, this configurational integral is equivalent to
that of the full polymer system.)  Taking the low-density limit for a
homogeneous system gives
\begin{equation}
\label{eq:v3}
- w^{(3)}(r_{12},r_{23},r_{13}) = \lim_{\rho \rightarrow 0}\ln \left[
\frac{g_3(r_{12},r_{23},r_{13})}{g_2({r}_{12})g_2({ r}_{23})g_2({
r}_{13})}\right]
\end{equation}
The $g_2({ r})$ ensure that the contributions due to the pair
interactions are subtracted from the triplet interaction
(c.f. Eq.~\ref{eqIII.2}). For our homogeneous system the 3-body
potential depends only on the three variables $\{ r_{12}, r_{23}, r_{13}\}
$ shown in Fig.~\ref{fig:3-polymers}. Even then, calculating the
triplet interaction $w^{(3)}(r_{12}, r_{23}, r_{13})$ for every
possible triplet arrangement is very cumbersome. We therefore confine
ourselves to configurations that make up an equilateral triangle.
Instead of three variables, the potential now depends only on the
length $r$ of each side of the triangle, simplifying the calculation
of Eq.~(\ref{eq:v3}) to:
\begin{equation}
w^{(3)}(r) = -\lim_{\rho \rightarrow 0} \left[ \ln g_3(r) - 3
w^{(2)}(r) \right],
\end{equation}
where we also used Eq.~(\ref{eqIII.3}).
  We expect that for $r=0$, i.e.\ complete overlap, the three body
interaction will be strongest, while for large $r$ the interaction
should vanish.

\begin{figure}
\begin{center}
\epsfig{figure=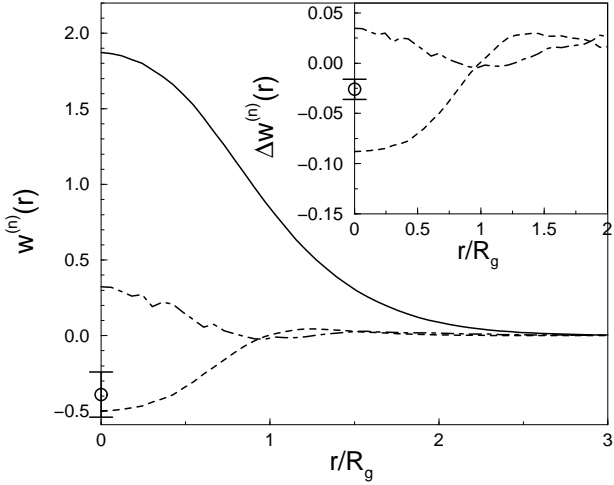,width=8cm}
\end{center}
\caption{\label{fig:v34} Effective potentials $w^{(n)}(r)$ for $n=2$
(solid line), $n=3$ (dashed line), $n=4$ (dash-dotted line), and $n=5$
(symbol with error bar at $r=0$).  Inset: relative potentials $\Delta
w^{(n)}(r)$ for $n=3$ (dashed line), $n=4$ (dash-dotted line), and
$n=5$ (symbol with error bar at $r=0$).}
\end{figure}

We calculated $w^{(3)}(r)$ for three $L=500$ SAW polymers on a
lattice. At this infinite dilution, $g_3(r)$ is simply the probability
that three polymers in a configuration $\{ r_{12}, r_{23}, r_{31} \}$
do not overlap.  In the Monte Carlo simulation we integrate over the
monomeric degrees of freedom by performing pivot moves. Once every
1000 MC steps we move the polymers into a triangular configuration and
check for overlap. The results are plotted in Fig.~\ref{fig:v34}.
Since the total free energy increases with the number of polymers, a
more relevant measure of the 3-body interactions is the relative
potential $\Delta w^{(3)}(r)= w^{(3)}(r)/3w^{(2)}(r)$ which denotes
the strength of the 3-body interaction relative to that  of the two
body interactions.  As shown in the inset of Fig.~\ref{fig:v34}, the
relative contribution of the 3-body potential to the total free energy
is quite small, only about $9\%$ of the contribution from the pair
potentials.

\subsection{4-body interactions}

Next we turn to the 4-body interactions.  Again, even for a
homogeneous phase, the total number of relevant coordinates makes the
calculation of the full interaction prohibitively complex.  To
restrict the number of coordinates in our calculations we determine
the 4-body potential by placing the four polymer CM's on a regular
tetrahedron and determining the non-overlap probability as a function
of the length $r$ of each side of the tetrahedron. The 4-body
potential is then defined as
\begin{equation}
 w^{(4)}(r) = -\lim_{\rho \rightarrow 0} \ln \left[ g_4(r) \right] - 4
w^{(3)}(r) - 6 w^{(2)}(r)
\end{equation}
since the tetrahedral configuration includes four equilateral
triangles (3-body interactions) and six edges (2-body
interactions).  The full and  relative 4-body interaction, $\Delta
w^{(4)}(r)= w^{(4)}(r)/ (4w^{(3)}(r) + 6w^{(2)}(r))$, are plotted in
Fig.~\ref{fig:v34}. Note that the 4-body interaction is smaller in
absolute magnitude and has the opposite sign to the 3-body
interaction. The relative contribution of the 4-body interactions to
the total free energy is less than $5\%$ of the total potential, and
also less than the relative 3-body contribution.

\subsection{5-body interactions}

Calculating the 5-body interaction is even more complicated than the
4-body interaction, and so we only evaluate it at full overlap -- when
all the CM's coincide -- where we expect its contribution to be
largest.  More generally, for any $n$th order term the interaction at
full overlap is given by:
\begin{equation}\label{eqIIIe1}
 w^{(n)}(0) = -\lim_{\rho \rightarrow 0}\left[\ln g_n(0) \right] -
\sum_{m=2}^{m=n-1} {n \choose m} w^{(m)}(0)
\end{equation}
where $\lim_{\rho \rightarrow 0} g_n(0)$ is the normalized probability
of full overlap of the CM of $n$ polymers. As long as the particles
are equidistant from each other, the same combinatorial expression
holds for finite $r$.  For the 5-body term we find that $w^{(5)}(0)=
-0.4 \pm 0.15 $, while the relative contribution of the 5-body terms
is given by $w^{(5)}(0)/(5 w^{(4)}(0) + 10 w^{(3)}(0) + 10 w^{(2)}(0))
= 0.027 \pm 0.01$.  Again, the relative contribution of the 5-body
term to the free energy is smaller and of opposite sign to those of
the 4-body terms.  Going beyond the 5-body interaction, even at
complete overlap becomes increasingly difficult.  For example, for the
5-body interaction, of the $10^8$ overlap checks, each attempted after
$1000$ pivot moves, only about $30$ resulted in non-overlap.  For the
6-body interaction we estimate that $10^{11}$ MC attempts would be
needed.  Another problem arises from finite monomer density.  As more
and more polymers overlap, the monomer density increases, so that in
practice for a given polymer length $L$, only a finite number of
multiple overlaps are possible.  We found previously that the largest
finite-size corrections to the scaling limit were at $r=0$ for 
$w^{(2)}(0)$\cite{Bolh01a}. The same probably holds for the higher
order interactions.  However, our $L=500$ calculations should still be
very near the scaling limit.

\section{Effective density dependent pair interactions}

From the previous section we see that explicitly calculating the {\em
density independent} interactions $w^{(n)}({\bf r}_{i_1},{\bf
r}_{i_2}\ldots {\bf r}_{i_N})$ becomes
rapidly more complex with increasing order $n$. Calculating all higher
order terms is therefore impossible.  In this section, we describe a
way to include the average effect of all higher order terms by
extending the relationship between the pair-interactions and the
pair-correlations to finite density $\rho$. This leads to a {\em
density dependent} effective pair interaction $v(r;\rho)$.

\subsection{Inverting pair-correlations to derive density dependent
pair potentials}

Although for finite densities there is no known direct functional
relationship of the type of Eq.~(\ref{eqIII.3}), there is a theorem
which states that for any given pair-correlation function $g_2(r)$ and
density $\rho$, there exists (up to an additive constant) a unique
pair potential $v(r;\rho)$ which exactly reproduces $g_2(r)$ at that
density\cite{Hend74,Reat86}.  If the original $g_2(r)$ is generated by
a system with only pair interactions, then $v(r;\rho) = w^{(2)}(r)$
will be independent of density.  If there are any higher order
interactions in the original system which influence the structure,
then this equivalence will only hold for $\lim_{\rho \rightarrow 0}
v(r;\rho) = w^{(2)}(r)$.  At finite densities $v(r;\rho)$ must change
since the structure is no longer equal to the one generated by
$w^{(2)}(r)$ alone.  Therefore $v(r;\rho)$ must be density dependent.

\begin{figure}
\begin{center}
\epsfig{figure=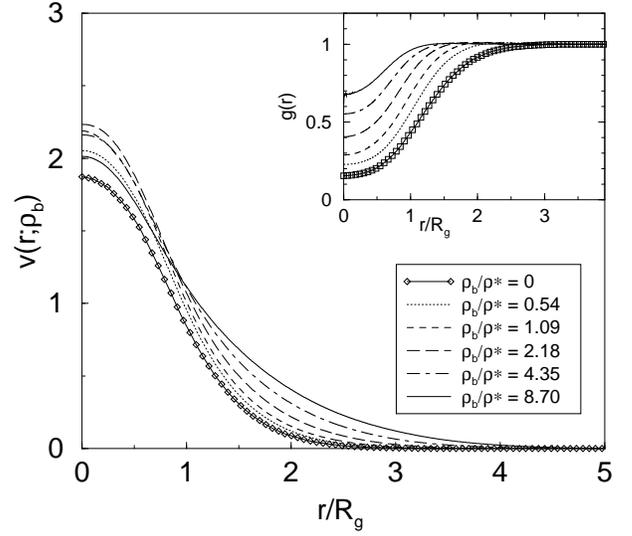,width=8cm}
\caption{\label{fig:veffL500} The effective polymer pair potentials
$v(r;\rho)$ derived at different densities from an HNC inversion of
the CM pair distribution functions $g_2(r)$ of polymer coils depicted
in the inset. (from Ref\protect\cite{Bolh01a})}
\end{center}
\end{figure}

\begin{figure}
\begin{center}
\epsfig{figure=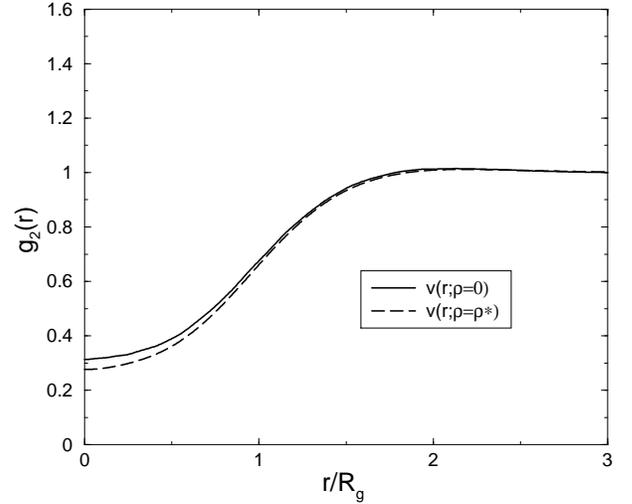,width=8cm}
\caption{\label{fig:gofr-rho0}
 Comparison of $g_2(r)$~'s generated at
density $\rho = \rho^*$ by the low-density potential $v(r;\rho=0)$ and
the correct potential $v(r;\rho=\rho^*)$.  Note how small the differences
are.
}
\end{center}
\end{figure}

In fact, this is what we found in two previous
papers\cite{Loui00a,Bolh01a}, where we used the hypernetted-chain
(HNC) approximation from liquid state theory\cite{Hans86} to extract
$v(r;\rho)$ from computer simulations of the $g_2(r)$'s between
the CM's of SAW polymers.  For completeness, we show these effective
pair interactions in Fig.~(\ref{fig:veffL500}). As expected, there is
a clear density dependence.  Without going into much detail about the
inversion of $v(r;\rho)$ from $g_2(r)$, we do want to point out that
the process can be very subtle.  As illustrated in
Fig.~\ref{fig:gofr-rho0}, the $g_2(r)$'s generated at $\rho=\rho^*$ by
$v(r;\rho=0)$ and $v(r;\rho=\rho^*)$ are very similar.  Any technique
to derive $v(r;\rho)$ from $g_2(r)$ must be significantly more
accurate than the difference between the $g_2(r)$'s shown in the
figure.  The accuracy of the techniques we use has been discussed in
Ref.~\cite{Bolh01a}, and will be analyzed in much more detail in
another publication\cite{Bolh01b}.

Any approximation which correctly reproduces the pair-correlations
will also predict the correct thermodynamics through the
compressibility equation\cite{Hans86,Volume}.  For our
density dependent $v(r;\rho)$ this is indeed the case, since we found
good agreement between the equation of state (EOS) $\Pi/\rho$
generated by the effective potentials in Fig.~\ref{fig:veffL500} and
the EOS of the underlying SAW polymer solution.  In contrast, the
$v(r;\rho=0)$ potential underestimates the EOS, and we find mean-field
fluid behavior\cite{Loui00a,Liko01b} $\Pi/\rho \sim \rho$ at large $\rho$ instead of the
correct $\Pi/\rho \sim \rho^{1.3}$ scaling.  So, even though the
$\rho=0$ potential results in pair-correlations $g_2(r)$ that are
similar to the true $g_2(r)$'s, the effective thermodynamics can
differ significantly.  The difference arises from the neglected
many-body interactions, as discussed  in
the Appendix.

\subsection{Understanding the density dependence of the effective
pair-potential}

Given the success of the density-dependent pair interaction in
describing pair-correlations and thermodynamics, we next turn to the
question of whether the density-dependence of $v(r;\rho)$ can be
directly understood from the density independent many-body
interactions.

Within the HNC approximation, the following expression due to Reatto
and Tau\cite{Reat87} and also Attard\cite{Atta92}
\begin{eqnarray}
\label{eq:axilrod}
v(r_{12};\rho) & = & w^{(2)}(r_{12}) - \rho \int \left(
e^{-w^{(3)}(r_{12},r_{13},r_{23})} -1 \right) \\ \nonumber & & \times
g_2(r_{13};\rho) g_2(r_{23};\rho) d{\bf r}_3,
\end{eqnarray}
describes the density dependence of the pair-potential that would
reproduce the true pair-correlations induced by the 2-body and 3-body
potentials. This is a generalization of earlier
expressions\cite{Rowl84,Rush67} and neglects terms of order ${\cal
O}(\rho^2)$ and higher.  In the literature it has mainly been applied
to the Axilrod-Teller interaction for rare-gas fluids, where it works
remarkably well, see e.g.\cite{Anta97} and references therein.

\begin{figure}
\begin{center}
\epsfig{figure=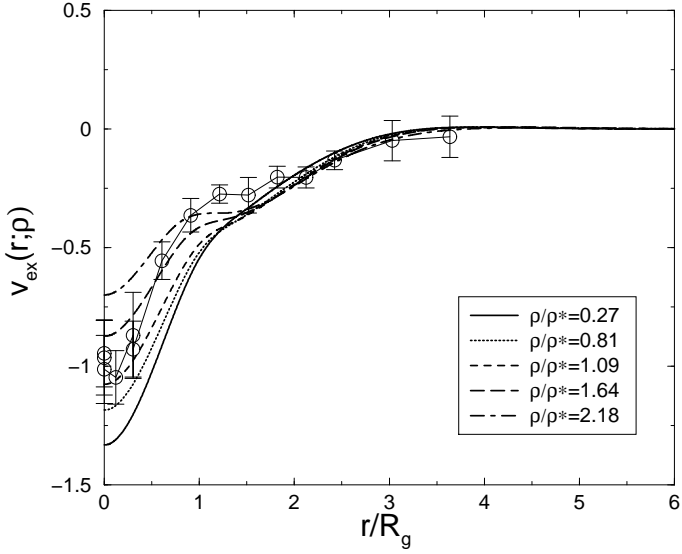,width=9cm}
\caption{\label{fig:densdep} The density dependent excess potentials
$v_{ex}(r;\rho)$ determined for several densities from the SAW
simulations are denoted by the smooth spline fit curves. The open
circles with error bars correspond to the evaluation of
Eq.~(\protect\ref{eq:vex}) at $\rho=0$, the line through them is to
guide the eye. }
\end{center}
\end{figure}

Figure~\ref{fig:densdep} highlights the density dependence by plotting
$v_{ex}(r;\rho)= (v(r;\rho) - w^{(2)}(r))/\rho$.  For clarity, we have
replaced the rather noisy data by spline fits.  For densities
$\rho/\rho^* <1$ the curves are close to each other suggesting that
the roughly linear density dependence in Eq.~(\ref{eq:axilrod}) holds
true.  For larger densities into the semi-dilute regime,
$v_{ex}(r;\rho)$ becomes smaller in magnitude and the density
dependence becomes non-linear.  This non-linearity is not unexpected,
since Eq.~(\ref{eq:axilrod}) neglects higher order terms in $\rho$, as
well as the effects of 4-body and higher order interaction terms.

We can go even further and directly calculate the triplet induced
density dependent term $v_{ex}(r;\rho)$ by substituting
Eq~(\ref{eq:v3}) into Eq.~(\ref{eq:axilrod}) to obtain
\begin{eqnarray}\label{eq:vex}
v_{ex}(r_{12},\rho) & = & - \int \left( \lim_{\rho \rightarrow 0}
\frac{g_3(r_{12},r_{13},r_{23})}{ g_2(r_{12}) g_2(r_{13}) g_2(r_{23})}
-1\right) \\ \nonumber & & \times g_2(r_{13};\rho) g_2(r_{23};\rho)
d{\bf r}_3.
\end{eqnarray}
The $g_2(r;\rho)$, in contrast to the $g_2(r)$ in the first term
in~(\ref{eq:vex}), are defined at the density of interest.  Evaluating
this integral is difficult, because the term between brackets can
become very small.  We use a direct MC procedure, where two polymer
coils are held with their CM a distance $r_{12}$ apart while we
integrate over the position of the third particle. In order to ensure
that the integral converges it is crucial to use the $g_2(r)$ at
$\rho=0$ (i.e.\ those between the brackets in Eq.~\ref{eq:vex}), from the
simulation itself, by calculating it on the fly. This is necessary to
avoid small errors in the radial distribution function which build up
during the integration over the volume. The $g_2(r;\rho)$ at finite
density are known from previous calculations. The resulting
$v_{ex}(r;\rho)$ for $\rho=0$ is plotted in
Fig.~\ref{fig:densdep}.  The results for finite density do not differ
by  very much.

In conclusion, the density dependence is mainly caused by 3-body
interactions, at least in the dilute regime.  At semi-dilute densities,
higher order many-body interactions may come into play.

\section{Many-body correlations}

If $g_2(r)$ is generated by a system with only pair-potentials, then
the exact inversion of $g_2(r)$ at {\em any} density will reproduce
the exact pair potential.  For such a system, the inverted pair
potential can be used in principle to determine all higher order
correlation functions of the original system.  In other words, for
systems with only pair-interactions, the pair-distribution function
$g_2(r)$ contains enough information to generate all higher order
correlation functions\cite{Evan90}!

If a system has 3-body or higher order interactions, then our
$v(r;\rho)$ still exactly reproduces the pair-correlations. But, as we
shall demonstrate in this section, it can no longer exactly reproduce
the higher order correlations.  Nevertheless, we will show that the
differences are not very large for the case of SAW polymers at the
densities we study.

\subsection{Bond angle distribution from 3-body correlations}

\begin{figure}
\begin{center}
\epsfig{figure=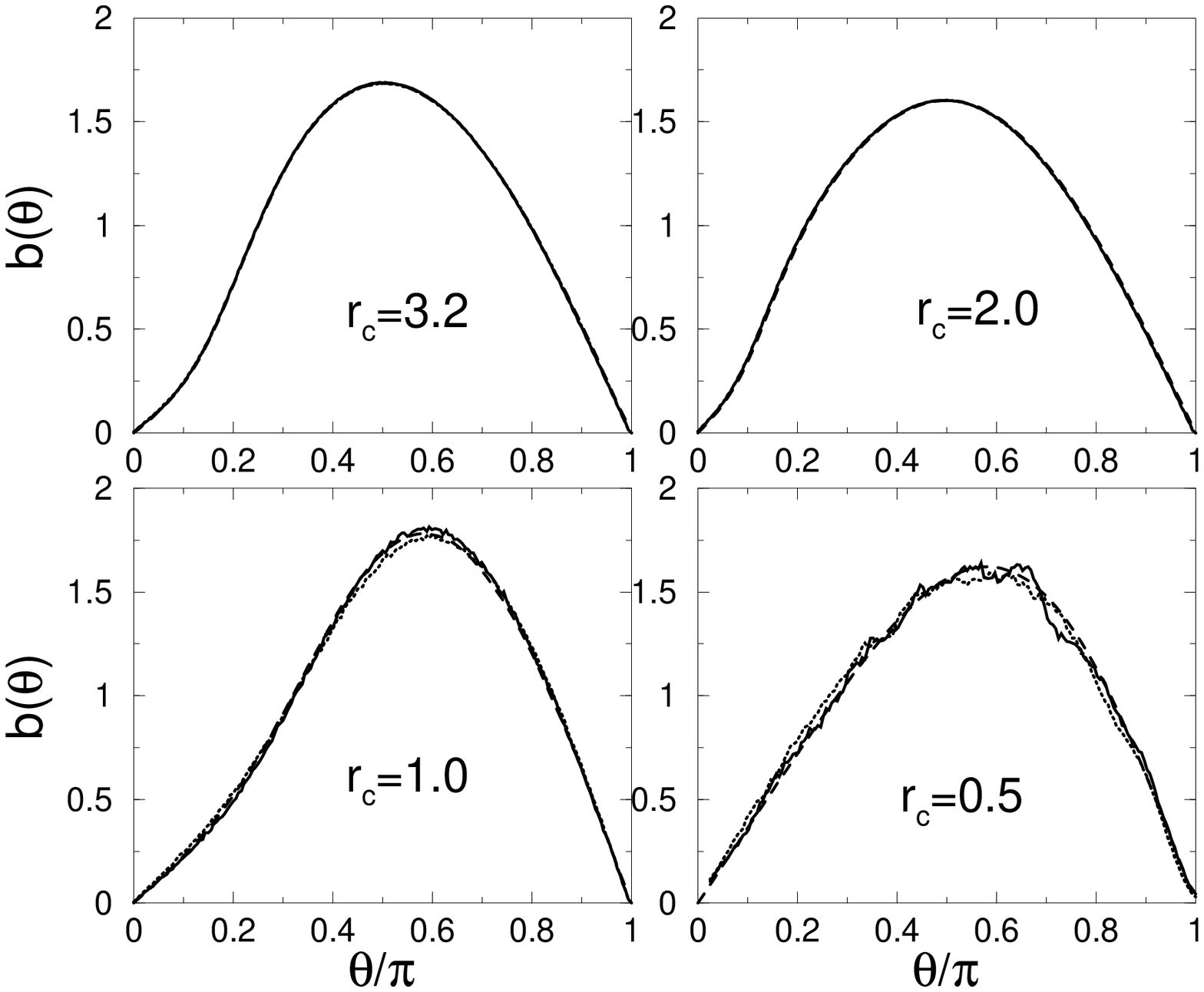,width=8cm}
\caption{\label{fig:bondangle} The bond angle distribution $b(\theta,
r_c)$ plotted for several cutoff radii $r_c$ for $\rho = \rho^*$. The
solid curves denote the effective potential results, the dotted curves
correspond to the explicit SAW simulations and the dashed curves show
Kirkwood's superposition approximation.  }
\end{center}
\end{figure}

\begin{figure}
\begin{center}
\epsfig{figure=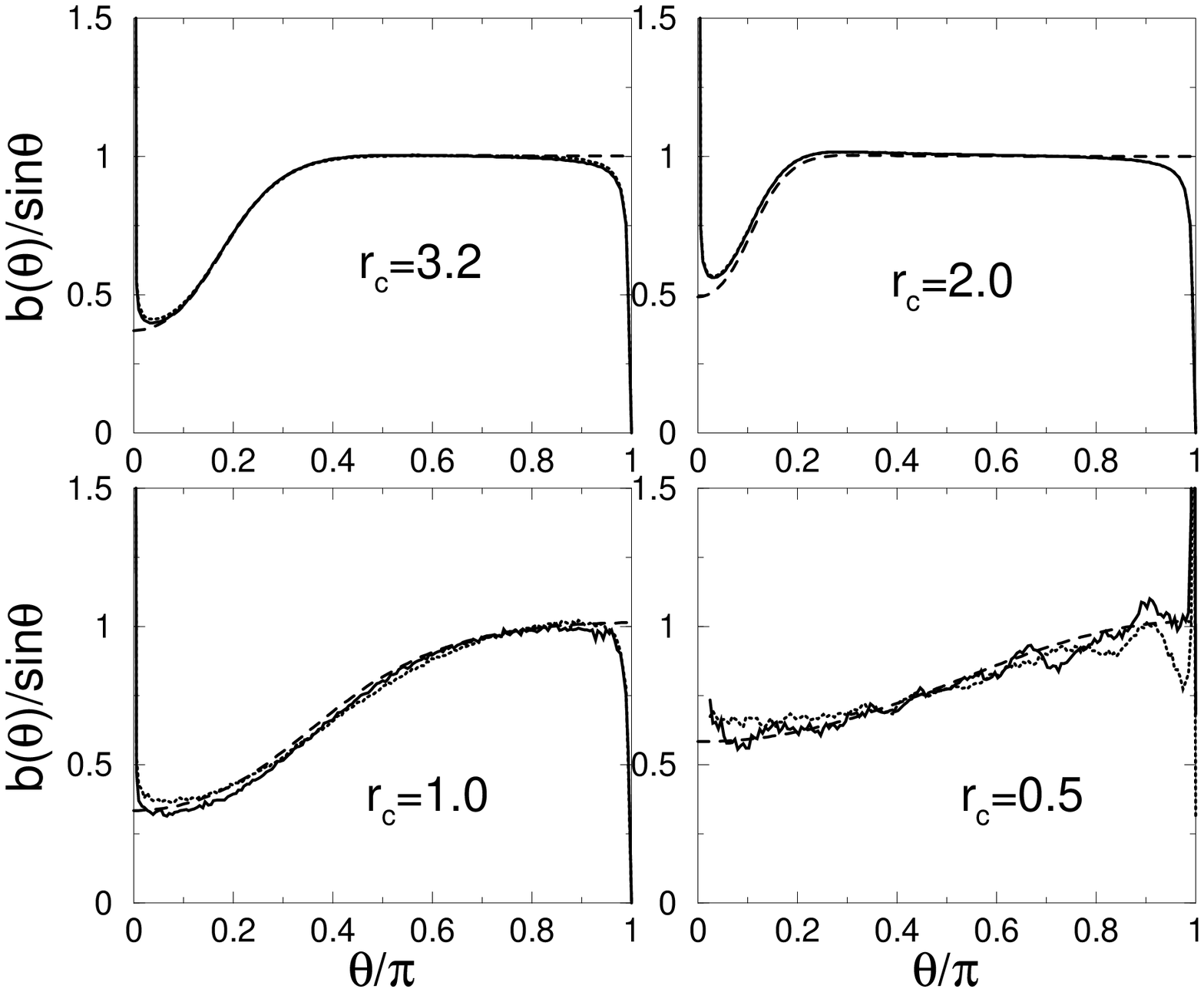,width=8cm}
\caption{\label{fig:bondangle_norm} The normalized bond angle
distribution $b(\theta, r_c)$ plotted for several cutoff radii $r_c$
for $\rho = \rho^*$. The solid curves denote the effective potential
results, the dotted curves correspond to the explicit SAW simulations
and the dashed curves show Kirkwood's superposition approximation. }
\end{center}
\end{figure}

Calculating and comparing the full 3-body correlation functions would
be very cumbersome for many of the same reasons that it is difficult
to map out the full 3-body interaction.  Therefore, we resort to a
reduced picture where a subset of the variables are integrated
out~\cite{Lowe93,Silva98}.  One popular measure of the 3-body
interactions is the bond angle distribution function, defined as
\begin{eqnarray}
\label{eq:bondorder}
b(\theta,r_c) &= & 8 \pi^2 \rho^2 N \int_0^{r_c} \int_0^{r_c}
g_3(r_{12},r_{13},(r_{12}^2 + r_{13}^2 \\ \nonumber & & -2 r_{12}
r_{13} cos \theta )^{1/2}) r_{12}^2 r_{13}^2 sin \theta
dr_{12}dr_{13},
\end{eqnarray}
where $N$ is a dimensionless normalization constant.  This integral
sums over all triplets within a cutoff radius $r_c$ from the central
particle and determines the distribution of the angle $\theta$ in
these triplets. We calculated the bond angle distribution for both the
SAW simulations and the effective pair potentials for different cutoff
radii $r_c$ as shown in Fig.~\ref{fig:bondangle}. The effective
potentials $v(r;\rho)$ reproduce this measure of the 3-body
correlations remarkably well.  Since for an ideal gas the bond angle
distribution exactly follows a sine curve, dividing the bond angle
distribution by $\sin \theta$ highlights the deviations from ideal
behavior. In Fig.~\ref{fig:bondangle_norm} we show the renormalized
bond angle distribution. The differences between the curves are now
clearer.  The absolute  deviations from the sine-like behavior
are largest at small $\theta$ because the particles repel each other
and triplets with small $\theta$ will be relatively rare. In the case
of a hard sphere systems this correlation hole would be even more
pronounced.  It is remarkable how well the bond order distribution
follows the ideal sine curve for the larger angles. At a cutoff radius
of $r_c=0.5$ the distribution becomes flatter, reflecting the 
broad flat top of the repulsive Gaussian shaped pair potential.
 
Instead of determining the distributions from explicit simulations, we
can also substitute the Kirkwood superposition
approximation\cite{Kirk52}
\begin{equation}\label{Kirkwood}
g_3(r_{12},r_{13},r_{23})\approx g_2(r_{12})g_2( r_{13})g_2(r_{23})
\end{equation}
into Eq.~\ref{eq:bondorder} and calculate the integral directly by
using the radial distribution functions from previous simulations.
This approximation is also included in Fig.~\ref{fig:bondangle} and
Fig.~\ref{fig:bondangle_norm} and turns out to be very accurate,
except for $\theta \approx 0$ and $\theta \approx \pi$ (see
Fig.~\ref{fig:bondangle_norm}) where the simulations are prone to
large statistical errors, due to the vanishing volume of the available
phase-space.

\begin{figure}
\begin{center}
\epsfig{figure=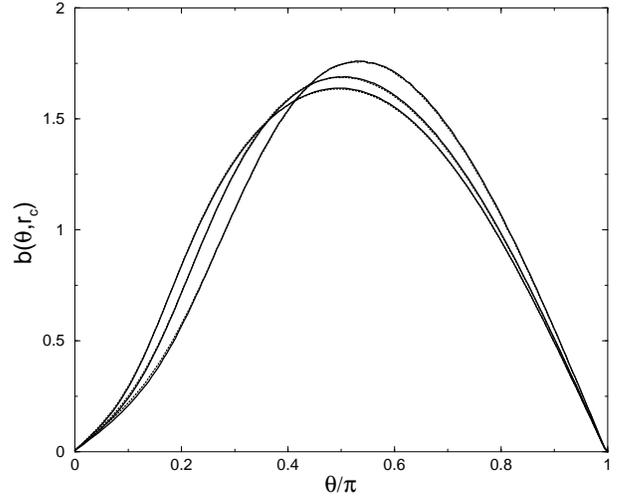,height=8cm,angle=-90}
\caption{\label{fig:notsens} Comparison between bond angle
distribution $b(\theta, r_c)$ for effective potentials $v(r;\rho=0)$
(dotted lines) and $v(r;\rho=\rho^*)$ (solid lines), both determined
from simulations at $\rho=\rho^*$. From left to right the curves
correspond to $r_c= 2.5$, $r_c= 2.0$ and $r_c= 1.5$. The differences
are so small that the lines are virtually indistinguishable on the
scale of the graph.}
\end{center}
\end{figure}

The bond-angle distribution is not very sensitive to differences in
the full 3-body correlations (see e.g. Ref.~\cite{Silva98}), partially
because it is an integrated quantity. An example of this is given
in Fig.~\ref{fig:notsens}, where the bond angle distributions of the
effective potentials $v(r;\rho=0)$ and $v(r;\rho=\rho^*)$ are compared
for the same density $\rho = \rho^*$.  Clearly, there is hardly any
difference between the distributions.

\subsection{Deviations from Kirkwood superposition for 3-body correlations}

  A more sensitive measure of triplet correlations is the deviation
from the Kirkwood superposition approximation, Eq.~(\ref{Kirkwood}),
which we define as:
\begin{equation}\label{kirkwood}
G_3( r_{12},r_{13},r_{23}) =\frac{g_3( r_{12},r_{13},r_{23}) }{
g_2(r_{12})g_2( r_{13})g_2( r_{23})}.
\end{equation}
Since our $v(r;\rho)$ exactly reproduces the $g_2(r)$, this expression
should highlight any differences between the true $g_3$ and the $g_3$
arising from our effective potential picture.  To simplify, we limit
ourselves for a given $r_{12}$ to triplet configurations for which $r
= r_{23} = r_{13}$ (i.e.\ isoceles triangles).

\begin{figure}
\begin{center}
\epsfig{figure=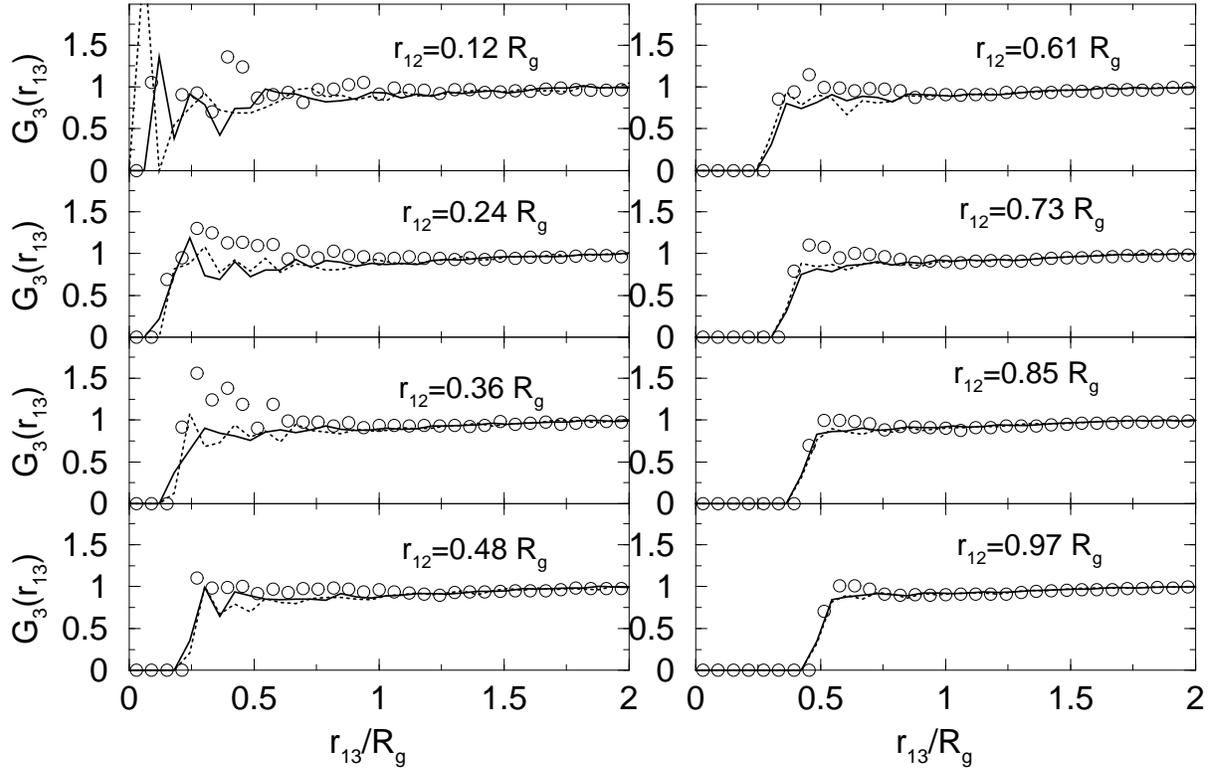,width=16cm}
\caption{\label{fig:g3_r0_r20all} To measure $G_3(r_{12},r_{13},r_{13})$,
which denotes deviations from Kirkwood superposition, we fix $r_{12}$,
and plot the $G_3(r)$ as  functions of $r=r_{13}=r_{23}$ (isoceles
triangles, see Fig.~\protect\ref{fig:3-polymers}).  Results are given
for SAW polymers (circles), effective potentials $v(r;\rho)$ (solid
lines), and $v(r;\rho=0)$ potentials (dashed lines). All plots are
from simulations at $\rho = \rho^*$.  Note that the two effective
potential plots are more or less identical within the statistical
noise, while the SAW $G_3^{SAW}(r)$ is slightly higher by approximately
a factor $\exp(-w^{(3)}(r))$.  }
\end{center}
\end{figure}

 First, we compare in Fig~\ref{fig:g3_r0_r20all} the $G_3(r)$ at
$\rho=\rho^*$ generated by $v(r;\rho=0)$ and by $v(r;\rho=\rho^*)$.
Just as we found for the bond angle distributions, the $G_3(r)$ are
very similar even though the potentials are different.  We already
showed that the $g_2(r)$ are not very different either (see
Fig.~\ref{fig:gofr-rho0}) , so that the same now holds for the full
$g_3(r_{12},r_{23},r_{13})$.

Next, we turn to a comparison between the true $G_3^{\rm SAW}(r)$
derived from explicit simulations of our SAW polymer system and the
$G_3^{\rm eff}(r)$ of the effective potentials at $\rho=\rho^*$ .  As
can be seen in Fig.~\ref{fig:g3_r0_r20all}, our effective pair-potential
$v(r;\rho)$ does not exactly reproduce the SAW 3-body correlation
function.  The trends are similar, but the deviation from
superposition of the SAW polymers, $G_3^{\rm SAW}(r)$, is consistently
larger than the same quantity generated by the effective potentials
$G_3^{\rm eff}(r)$, especially if $r < R_g$.

For systems with an explicit three-body interaction the Kirkwood
superposition approximation is sometimes written as:
\begin{eqnarray}
\label{kirkwood2}
g_3( r_{12},r_{13},r_{23}) & \approx & g_2(r_{12})g_2( r_{13})g_2(
r_{23}) \\ \nonumber & & \times \exp\left[
-w^{(3)}(r_{12},r_{13},r_{23})\right].
\end{eqnarray}
This is exact in the $\rho \rightarrow 0$ limit, as can be seen from
Eq.~(\ref{eq:v3}).  Note that in this same limit the three-body
correlations induced by the $v(r;\rho)$ reduce to the simpler Kirkwood
superposition approximation $g_3( r_{12},r_{13},r_{23})\approx
g_2(r_{12})g_2( r_{13})g_2( r_{23})$, demonstrating explicitly that in
contrast to the 2-body correlations, $v(r;\rho)$ cannot exactly
reproduce the 3-body correlations if there is a 3-body interaction
present!  In fact, we have shown explicitly that two systems with
identical pair correlations, namely our effective potential system and
the original SAW system, can have differing triplet correlations.

These arguments also suggest that a simple approximation, namely
$G_3^{SAW}(r) \approx G_3^{\rm eff}(r) \exp[-w^{(3)}(r)]$, can shed some
light on the differences observed in Fig~\ref{fig:g3_r0_r20all}.
Since $w^{(3)}(r)$ is negative for equilateral triangle
configurations, as illustrated in Fig~\ref{fig:v34}, it is perhaps not
surprising that roughly speaking $G_3^{SAW}(r) > G_3^{eff}(r)$ for the
isoceles triangle configurations plotted in
Figs~\ref{fig:g3_r0_r20all}, at least in the region where $w^{(3)}(r)$
is non-zero.  Unfortunately the statistical errors in this region are
very large, making a quantitative comparison difficult, but the
deviation is certainly of the same order as would be expected from an
extra factor $\exp[-w^{(3)}(r)]$ (compare with Fig.\ref{fig:v34}).

Our very simple approximation illustrates how deviations from Kirkwood
superposition originate from two effects: {\bf (1)} Deviations induced
by correlations generated by the pair-potentials alone.  These have
been studied in great detail for hard-sphere systems, see
e.g.\cite{Barr88,Khei99}. {\bf (2)} Deviations induced primarily by
3-body potentials.  In practice, of course, these two effects are
somewhat entwined, especially at higher densities.  Nevertheless
splitting the two effects can shed light on the origin of three-body
correlations.  In particular, it suggests that while an effective
pair-potential $v(r;\rho)$ that exactly reproduces the $g_2(r)$ can
partially reproduce deviations from superposition of type {\bf (1)},
it will fail for deviations of type {\bf (2)}.

Since the 3-body and higher order interactions between the CM of
polymer solutions are not very strong, the total $g_3(r)$'s are still
remarkably well reproduced by the $v(r;\rho)$, especially when
integrated quantities such as the bond-angle correlations are
considered.  However, for systems where 3-body interactions are
strong, such as liquid Si or liquid Ga, one cannot expect the same
success from effective pair potentials.  Very similar conclusions were
stressed by Evans\cite{Evan90} in the context of reverse Monte-Carlo
simulations\cite{Silva98,Mcgr92}.

\section{conclusions}

Integrating out the monomeric degrees of freedom to obtain a
description based on effective potentials between  polymer CM's is a
useful coarse-graining technique for polymer solutions.  Because
simulations can be performed to high accuracy, the lessons learned
here should be applicable to a much broader range of coarse-graining
schemes.

In particular, we showed that the free energy of the polymers can be
expanded in a series of  state independent many-body effective potentials.
The terms in the series oscillate as $(-1)^n$, and become smaller in
absolute magnitude for increasing $n$. The scaling theory developed
in the Appendix confirms these ideas, and can be  used to extend them
to arbitrary order $n$.

A parallel description of the coarse-grained polymer solution was
developed in terms of an effective state (density) dependent
pair potential $v(r;\rho)$, which exactly reproduces the
pair-correlations and, in an average way, includes all the higher
order terms in the many-body free energy expansion.  The
density-dependence of this effective pair-potential can be largely
understood from the direct influence of the density-independent
three-body interactions.

The three-body correlations are also well described by this effective
pair-potential picture.  If the bond angle distribution is used, the
differences between the full SAW polymer system and our effective
pair-potential picture are almost indistinguishable.  However, this is
not a good ``order parameter'' for measuring deviations in three-body
distribution functions, since there is almost no difference between
results from the full simulations and those produced with the Kirkwood
superposition approximation.  When we use a more direct measure of the
deviations from Kirkwood superposition, small differences between the
effective two-body and the full SAW triplet distributions can be
measured.  These arise primarily from the direct effect of the
three-body potential $w^{(3)}(r)$, and illustrate a more general
point, namely that an effective two-body interaction can never
simultaneously reproduce the two and three-body correlations in a
system with many-body interactions exactly.  Since for polymers in a
good solvent these many-body interactions are relatively weak, a
coarse-grained description based on effective pair-interactions works
remarkably well, at least for the dilute and the beginning of the
semi-dilute regime\cite{Loui00,Bolh01a}.  Whether this success can be
extended deeper into the semi-dilute or into the melt-regime, or even
for polymers in poor solvents, remains to be seen, and will be
the subject of future investigations.

\acknowledgements

 AAL acknowledges support from the Isaac Newton Trust, Cambridge, PB
acknowledges support from the EPSRC under grant number GR$/$M88839, We
thank H. L\"{o}wen and C. von Ferber for useful discussions, and
R. Finken and V. Krakoviak for a critical reading of the manuscript.

\appendix

\section{Relative strength of $n$ body terms from  scaling theory}

In a recent paper, von Ferber {\em et al.}~\cite{vonF00} used scaling
theory and simulations to calculate the triplet interaction for
star-polymers and found an attractive interaction with a relative
strength of about $11 \%$, very similar to our 3-body results.  The
natural choice for the position coordinate of a star-polymer is not the
CM, but its midpoint\cite{Liko98}. However, we will see that for estimating
relative contributions, the difference between our CM representation
and the midpoint representation is not that important.

Here, we apply the star-polymer scaling theory to estimate the
relative contributions of $w^{n}(0)$ to all orders in $n$. We
specialize to linear polymers, which can be seen as star-polymers with
only two arms.  We first note that the partition function for $n$
polymers with their mid-points constrained to be at a distance
$r \ll R_g$ apart scales as\cite{vonF00,Dupl89,vonF97}:
\begin{equation}\label{eqIIIf1}
Z_n(r) \sim r^{\theta_n},
\end{equation}
in the limit $r/R_g \rightarrow 0$. Here $\theta_n$ is the contact
exponent which in turn can be written as
\begin{equation}\label{eqIIIf2}
\theta_n = n \eta_2 - \eta_{2\cdot n}.
\end{equation}
where the $\eta_f$ are the scaling exponents for a star-polymer with
$f$ arms. These are tabulated for two different renormalization group
calculations in\cite{vonF00,vonF97b}, and can also be approximated by
a simpler expression
\begin{equation}\label{eqIIIf7}
\eta_f \approx 0.335307 f^{3/2}
\end{equation}
which is expected to become more accurate for larger
$f$\cite{Ohno89}.  Comparing the different approximations gives
an indication of the accuracy of the scaling theory results.

The probability of finding $n$ polymers with their midpoints a
distance $r \ll R_g$ apart can be found from the partition functions
since:
\begin{equation}\label{eqIIIf3}
-\lim_{\rho \rightarrow 0} \ln g_n^{mid}(r) \propto \ln \left[
 Z_n(r)\right] \sim \theta_n \ln (r/R_g).
\end{equation}
Here $g_n^{mid}(r)$ is the $n$-body distribution function for the
midpoints of $n$ polymers.  These scaling theories can be developed
for polymers in the midpoint representation in part because of the
analogy of the overlap of $n$ polymers to a $2n$ arm star-polymer.
Such an analogy is not as obvious for the CM representation, hampering
the derivation of a CM based scaling theory.  On the other hand, MC
simulations for overlap in the CM representation are much easier than
for the midpoint representation because the probability $g_n(r)$ is
much larger than $g_n^{mid}(r)$ in the small $r$ limit, in fact
$g_n^{mid}(0)=0$ while the CM $g_n(0)$ is finite. Gathering good
statistics for the mid-point representation is thus much slower than
for the CM representation.  It nevertheless seems reasonable to expect
that the {\em relative} strengths of different orders of the
interactions in the CM representation are similar to those of the
midpoint representation, suggesting that the relative probability of
full overlap of the CM of $n$ polymers scales as:
\begin{equation}\label{eqIIIf4}
\lim_{\rho \rightarrow 0} \left(\frac{\ln g_n(0)}{ \ln g_m(0)}\right)
\approx \lim_{r \rightarrow 0}\lim_{\rho \rightarrow 0}\left(
\frac{\ln g_n^{mid}(r)}{\ln g_m^{mid}(r)} \right) \approx
\frac{\theta_n}{\theta_m}.
\end{equation}
In Table~\ref{tableIII} we confirm this ansatz by comparing $\ln
g_n(0)/ \ln g_2(0)$ calculated by direct simulations for $n=3,4,5$
with $\theta_n/\theta_2$ calculated by three different versions of the
scaling theory.  Clearly, all three scaling theories and the
simulations agree reasonably well with each other, giving us
confidence to proceed.

Armed with this approximate equivalence of the $g_n$'s in the two
representations, we can now fruitfully compare the strength of the $n$
body interaction $w^{(n)}(0)$ for CM particles to an expression
derived from the scaling theory, by using Eq.~(\ref{eqIIIf2}) and
Eq.~(\ref{eqIIIf3}) to recursively rewrite Eq.~(\ref{eqIIIe1}) as:
\begin{equation}\label{eqIIIf5}
 w^{(n)}(0) \propto \hat{w}^{(n)} = (-1)^n \sum_{m=2}^{n} { n \choose
m} (-1)^{m} (m \eta_2 - \eta_{2 \cdot m}).
\end{equation}
Here $\hat{w}^{(n)}$ is the coefficient of the full (divergent)
midpoint-midpoint interaction $-w_{mid}^{(n)}(r)~\sim~\hat{w}^{(n)}
\ln(r/Rg)$ in the limit $r \rightarrow 0$. The values of these terms
relative to the $n=2$ term are compared in Table~\ref{tableIII} with
direct simulation results for $n=3,4,5$, and again good agreement is
obtained.  For the scaling theory each higher order term is opposite
in sign and smaller than the previous one.  Our simulations show the
same trend for the sign, but the error bars on the $n=5$ simulations
are still too large to ascertain that its magnitude is less than the
$n=4$ term.

The strength of the $n$ body term contribution to the free energy
relative to all lower order terms can also be found from the arguments
above. In this case the pre-factors cancel and we find:
\begin{equation}\label{eqIIIf6}
\Delta w^{(n)}(0) \approx \Delta \hat{w}^{(n)}=
\frac{\hat{w}^{(n)}}{\theta_n - \hat{w}^{(n)}}
\end{equation}
Once again the simulations and the scaling theory agree quite well
both for the magnitude and the sign of the different terms.

\section{Convergence of the full free energy series}\label{Convergence}

The good agreement found between the scaling theory and the
simulations for $n=3,4,5$ suggests that we can evaluate trends for
arbitrary order $n$ from the scaling theory. We use the simple
expression in Eq.~(\ref{eqIIIf7}) which gives $\eta_f$ for arbitrary
$f$, to simplify Eq.~(\ref{eqIIIf5}) to
\begin{equation}\label{eqIIIf8}
\hat{w}^{(n)} = 0.9484 (-1)^{n} \left(\sum_{m=1}^{n} { n \choose m}
(-1)^{m+1}m^{3/2}\right).
\end{equation} 
The sum between brackets is positive for all $n$, and becomes smaller
for each subsequent term.  For example for $n=3$ it is $0.29$, while
for $n=10,000$, it is has dropped to $0.009$.  This implies that at
all orders, the $n$-body interaction $w^{(n)}(0)$ oscillates in sign
as $(-1)^n$ and decreases in absolute magnitude with increasing $n$.
The absolute magnitudes of the relative contributions $\Delta
w^{(n)}(0)$ decrease even faster with $n$.

We expect that the absolute magnitude of each $n$ body term is largest
at full overlap.  However, the fact that each higher order term
decreases in absolute magnitude, or even that the terms oscillate in
sign at $r=0$, does not necessarily imply that the full free energy
expansion of Eq.~(\ref{eqIII.1}) will converge quickly for all
configurations $\{{\bf r}_i\}$ of the CM.  For one thing, for an $N$
particle system, the number of $n$-tuple coordinates grows as the
binomial $N!/(N-n)!n!$. Therefore, even though the magnitude of the
$n$ body terms $w^{(n)}({\bf r}_{i_1},{\bf r}_{i_2}\ldots {\bf
r}_{i_n})$ becomes smaller with increasing $n$, the number of such
terms at each order $n$ in the free energy series Eq.~(\ref{eqIII.1})
increases almost exponentially with increasing $n$.  Of course for a
given configuration $\{{\bf r}_i\}$, not only the magnitude, but also
the range of each term $w^{(n)}({\bf r}_{i_1},{\bf r}_{i_2}\ldots {\bf
r}_{i_n})$ is needed to decide how many, if not all $N$ terms, are
needed to calculate the full free energy to a desired accuracy.  As
mentioned before, calculating the complete dependence of $w^{(n)}({\bf
r}_{i_1},{\bf r}_{i_2}\ldots {\bf r}_{i_n})$ on all possible $n$-tuple
CM coordinates $ \{ {\bf r}_{i_1},{\bf r}_{i_2}\ldots {\bf r}_{i_n}
\}$ is usually an impossible task for higher order $n$.

Nevertheless, some progress can be made by looking at the special case
of a set $\{{\bf r}_i\}$ where all $N$ CM's are at the same point.
This is simply the situation studied in the previous section by
scaling theory, for which the free energy is given by
\begin{equation}\label{eqIIIf9}
{\cal F}(N,V,\{{\bf r}_i\}) - {\cal F}^{(0)}(N,V) = -\ln g^N(0) \propto
\theta_N.
\end{equation}
Ignoring for the moment the volume term, who's contribution is
expected to be negligible\cite{Liko01,Bolh01a} we can ask the
question: How well is the free energy described by taking only
pair-interactions terms into account?  Since there are $\frac{1}{2}
N(N-1)$ pair terms, the free energy with only pair-terms taken into
account scales as ${\cal F} \propto N(N-1) w^{(2)}(0) \sim N^2$ while
the true free energy scales as ${\cal F} \propto \theta_N \sim
N^{3/2}$.  Including only the pair terms heavily {\em overestimates}
the free energy, a result also found by von Ferber {\em et
al.}\cite{vonF00}. In other words, if one truncates the coarse-grained
free energy series in Eq.~(\ref{eqIII.1}) then certain configurations
$\{{\bf r}_i\}$ will not be properly accounted for.

  However, if one were to use this coarse-grained free energy series
to study a given polymer solution, then the relative probability for
finding a configuration at complete overlap of the CM would be very
small. On the other hand, for any given order $n$, the probability of
finding $n$-fold overlap increases with increasing density, so that
one would expect to need more and more terms in the coarse-grained
free energy series as the density increases.

 Another way to measure the effects of truncating the coarse-grained
free energy series in Eq.~(\ref{eqIII.1}) would be to compare it to the
excess free energy per particle for the original polymer solution.
This can be found by averaging ${\cal F}(N,V,\{ {\bf r}_i \})$ over
all configurations $\{{\bf r}_i\}$ to obtain the total free energy
$F(N,V)$, as done in Eq.(\ref{eqIII.1b}). Including only pair
interactions in our CM representation of polymer solutions casts the
problem into that of finding the free energy of a mean-field
fluid\cite{Loui00a,Loui01a,Liko01b} for which the excess free energy
per polymer scales as:
\begin{equation}\label{eqIIIf10}
\frac{F^{ex}}{N} \sim \rho
\end{equation}
while the true excess free energy of a polymer solution in the
semi-dilute regime follows from scaling theory\cite{deGe79}
\begin{equation}\label{eqIIIf11}
 \frac{F^{ex}}{N} \sim \rho^{1/(3 \nu -1)} \approx \rho^{1.3}.
\end{equation}
Whereas including only the pair interactions in an average over all
configurations $\{{\bf r}_i\}$ in the semi-dilute regime {\em
underestimates} the total free energy, taking into account only pair
terms for single configuration $\{{\bf r}_i\}$ at complete overlap of
the CM {\em overestimates} the coarse-grained free energy.  This
implies that the special configurations where the coarse-grained free
energy series in Eq.~(\ref{eqIII.1}) breaks down do not have a strong
influence on the thermodynamics.

\end{multicols}
\newpage

\begin{table}
\caption{\label{tableIII} Comparison of scaling theory and simulations
for many-body interactions between the polymer CM. The labels a and b
denote results that follow from two different renormalization group
calculations for the exponents $\eta_n$\protect\cite{vonF00}, while label c comes from the
simple expansion of Eq.\protect\ref{eqIIIf7}, and label d denotes
simulation results for $L=500$ SAW polymer simulations.
}
\begin{tabular}[t]{llllllllll}
 $n$ & $2$ & $3$ & $4$ & $5$ & $6$ & $7$ & $8$ & $9$ & $10$ \\ 
\hline
a $\theta_n/\theta_2 $ & 1($\theta_2=0.8)$ & 2.55 & 4.49 & 6.75 & 9.3
& & & & \\ 
b & 1 ($\theta_2=0.82$) & 2.65 & 4.80 &7.41 & 10.43 & & & &
\\ 
c & 1 ($\theta_2=0.79)$ & 2.65 & 4.83 & 7.46 & 10.50 & 13.90 & 17.66
& 21.73 & 26.10 \\ 
d $\ln g_n(0)/\ln g_2(0)$ &1 ($\ln g_2(0)=1.88$) &
2.74 $\pm$ 0.01 & 5.13 $\pm$ 0.02 & 7.99 $\pm$ 0.05 & & & & & \\ 
\hline
 a $\hat{w}^{(n)}/\hat{w}^{(2)}$
&1 ($\hat{w}^{(2)}=0.8$)&-0.45 &0.29 & -0.19 & 0.11 & & & & \\ 
b &1
($\hat{w}^{(2)}=0.82$)&-0.35 & 0.22 & -0.15 & 0.085 & & & & \\ 
c &1
($\hat{w}^{(2)}=0.79$)&-0.35 & 0.22 & -0.17 & 0.14 & -0.12 & 0.11 &
-0.10 & 0.094 \\ 
d $w^{(n)}(0)/w^{(2)}(0)$ &1 ($w^{(2)}(0)=1.88$) &
-0.26 $\pm$ 0.013& 0.17 $\pm$ 0.031 & -0.21 $\pm$ 0.08& & & & &\\ 
\hline a $\Delta
w^{n}(0)$ & &-0.15 & 0.069 & -0.027 & 0.012 & & & & \\ b & &-0.12 &
0.048 & -0.019 & 0.0083 & & & & \\ 
c &&-0.12?? & 0.049 & -0.023 &
0.014 & -0.0088 & 0.0063 & -0.0046 & 0.0036 \\ 
d & & -0.09 $\pm$ 0.004 &
0.034 $\pm$ 0.007 & -0.026 $\pm$ 0.01 & & & & &
\end{tabular}
\end{table}

\newpage
\begin{multicols}{2}

\end{multicols}
\newpage

\end{document}